\documentstyle[prb,aps,preprint]{revtex}

\begin{document}
\draft
\title{Influence of Boundary Structure on a Light Absorption 
in Semiconductors} 
\author{L.~S.~Braginsky\thanks{e-mail: brag@isp.nsc.ru}}  
\address{Institute of Semiconductor Physics, 630090 
Novosibirsk, Russia} 
\date{April 30, 1997}
\maketitle
\begin{abstract}
The phenomenological boundary conditions for the envelope 
wave function, which is applicable for contacts of semiconductors 
with a rather different crystal symmetry are proposed.  
 It is shown that the boundary 
conditions are determined by the number of the real values, which are 
independent of the electron energy. The number of these 
parameters depends on the symmetry of the bordered materials as 
well as the symmetry of the boundary itself.

The proposed boundary conditions are used for the investigation 
of the light absorption at the indirect-band-gap semiconductor 
surface. It is shown that the possibility of the electron 
transitions with the momentum nonconservation could result in 
 enhancement of the absorption. This is especially the case for 
the small crystallites, which size is about 50\,\AA, and where 
the share of the surface atoms is sufficiently large.   The 
influence of the crystallite size as well as the structure 
 of the interface on the absorption are investigated.
\end{abstract}
\narrowtext
\section{ Introduction}

The problem of the boundary conditions for the envelope wave 
function is widely discussed (see, for instance, \cite{Ando,Burt} 
and references therein). The main point of the problem 
is as follows.  The 
electron behavior in the crystal subjected to the smooth 
field is usually described by means of the envelope 
wave function. This is more convenient then the Bloch wave function. 
The problem of their proper matching at the interface arises 
when two materials comes to the close contact. The boundary 
conditions where the wave functions and their derivatives 
supposed to be equal at the interface are in common usage in the 
quantum mechanics. However, this procedure fails if the 
effective masses of the contacted materials are different, 
because it leads to nonconservation of the probability flux 
through the boundary.  The most simple way to get around this 
difficulty has been proposed by Bastard \cite{Bastard}. He 
has supposed that the envelopes and their derivatives divided to the 
corresponding effective mass should be matched at the interface. 
It is shown that these boundary conditions hold at the smooth 
interface \cite{PR94,Burt}.  Nevertheless, they proved to be 
unacceptable in the simple models of the sharp boundary 
\cite{Zhu,FTT95}.

The problem becomes more complicated if the degeneration between 
the electron states from the different bands or valleys is 
present. In this case the intervalley conversion becomes 
possible at the interface. This problem has been considered 
numerically for the heterojunctions composed of the 
A$_{\rm III}$B$_{\rm V}$ 
semiconductors (GaAs/AlAs, HgTe/CdTe, GaSb/InAs) \cite{Ando}. It 
is shown, that the Bastard boundary conditions are acceptable, 
if they relate  the envelopes corresponded to the equivalent 
valleys from both sides of the contact. The parameters of 
the intervalley conversion were found to be small in these 
heterojunctions.  Nevertheless, the physical reason  for the 
Bastard boundary condition  to be applicable in this case is not 
clear yet. It seems, that the symmetry equivalence of both 
contacted crystals is important. This makes possible to 
classify the wave functions in each  material in the 
same way. The fact that both contacted materials are close 
chemically might be significant too.

Considerable recent attention has been focussed on the contacts 
of the rather different materials. First and foremost it 
 concerns the  surface of the many-valley semiconductors. The 
 photoemission from the GaAs surface into vacuum has been 
 considered in Ref.\cite{Lampert}.  The 
 quantum yield of the process was found to be quite less then 
 anticipated. Besides, the energy distribution threshold of the 
 emitted electrons suffered considerable downshift. It was 
 shown, that both  effects could be connected with the 
 peculiarities of the boundary conditions in this case 
\cite{FTT95,FTT97}.

 Currently considerably study is being given to the contacts of 
materials, which are rather different  on the crystalline 
symmetry as well as  on their chemical composition. This is 
especially true in regard to interfaces between the metal oxide 
semiconductors (TiO$_2$, Al$_2$O$_3$, ZrO$_2$) and the 
organic compounds \cite{Bignozzi}. These structures are used 
for the solar energy cells. Because of this, the charge transfer 
from the organic compound to the conductivity band of the 
semiconductor is of main important. The metal oxide layers are 
composed of interconnected mesoscopic particles. 
That is why the study  
 of the charge transfer between the 
nanocrystals as well as their optical properties are 
vigorously proceeding \cite{Nuesch,Kavan}.

The modern technology allows to prepare the quantum dots, where 
the small semiconductor crystallites are arranged in the certain 
matrix. The latter could be crystalline \cite{21} or amorphous 
\cite{13-20}, organic \cite{22} or nonorganic \cite{21}. Their 
optical spectra was found to be  quite sensitive to the 
structure of the boundary between the matrix and the crystallite 
\cite{120}.

It seems improbable, that Bastard boundary conditions are 
applicable to these contacts. Besides, the simple boundary 
conditions do not allow to describe the intervalley 
conversion of the electrons at the interface.  
The boundary conditions, which are appropriate for the contacts 
of the materials with  different crystal symmetry have been 
proposed in Ref.\cite{Laikhman}.  However, the $k-p$ 
approximation adopted in this paper restrict its applicability 
to the semiconductor contacts with the small band offsets. 
The contact two-valley crystal--one valley crystal has been 
investigated in Ref.\cite{FTT97}, where the simple 
one-dimensional microscopeing  model has been used to obtain the 
exact boundary conditions.

In this paper the phenomenological boundary conditions, which 
are applicable to the contacts of the material with the rather 
different band structures, are proposed. It is shown that the 
boundary conditions are determined by the certain array of the 
parameters. These parameters depend only on the contacted 
materials and on the technology of the contact preparing. They 
could be numerically calculated or measured in experiments. The 
magnitude of these parameters is estimated. 

The proposed boundary 
conditions are used to investigate the light absorption at the 
surface of the indirect-band-gap semiconductors.
It is well known that the light absorption near its edge in the 
indirect-band-gap semiconductors is much weaker than that in the 
direct-band-gap semiconductors. It happens because of the 
momentum conservation law. In order to satisfy it the third 
object has to be involved. It should carry out the extra momentum, 
which the electron obtains when it moves from the top of the 
valence band to the bottom of the conductive band.  This momentum 
is large (of the order of $\pi \hbar /a$, where $a$ is the 
lattice constant) in comparison with the photon momentum.  This 
third object may be either an impurity (it is important in the 
heavy doped semiconductors) or a short-wavelength phonon.  
The weakness of the interaction in both cases manifest itself in 
the additional small parameter, which has to appear in the 
theory.

It is clear that the momentum should not be conserved, if the 
absorption takes place at the sharp interface or 
at the surface. The surface becomes the above-mentioned third 
object in this case, and the proximity to it  determines the 
small parameter araised in the problem. The ratio $a/L$ (or 
more precisely, the certain degree of this ratio), where $L$ is 
the absorption length is the parameter, which specifies the 
proximity.  This value is too small when the light absorption in 
the bulk material is investigated.  Nevertheless, it is not the 
case for the restricted semiconductors.  Then $L$ becomes the 
size of the semiconductor, so that the constant of the 
electron-phonon interaction may become comparable or even less 
then this parameter. This means, that the 
considerable increase of absorption could be observed in the 
small semiconductor object.  The latter may be the quantum dots, 
wires, colloidal solutions of semiconductors, semiconductor 
polycrystals, or any other objects where the share of the 
surface atoms is sufficiently large.

This paper is organized as follows. In Sec.\,II 
the phenomenological boundary conditions for the envelope 
wave function  are derived.  The expressions for the parameters 
of the boundary conditions are presented and analyzed. In 
Sec.\,III the light absorption at the sharp semiconductor 
boundary is investigated. It is shown that its value as well 
as its behavior at the transparency edge is determined on the 
parameters of the boundary conditions. The results are discussed 
in Sec.\,IV and summarized in Sec.\,V. The units there $\hbar=1$ are 
adopted.

\section{Boundary conditions for envelope wave functions at 
sharp interface of materials with a different crystal symmetry}
The influence of the smooth external field on the electron in the 
crystalline lattice could be investigated in terms of the envelope 
wave function.
  In the one band approximation  the envelope  obeys the equation 
\begin{equation}
\label{1}
[\varepsilon_n(-i\nabla)+V({\bf r})]F({\bf r})=EF({\bf r}).
\end{equation}
Here  $\varepsilon_n(k)$  is the band spectra of an electron in 
the $n$-th band, $V({\bf r})$ is an external field, and F({\bf 
 r}) is the envelope. It is assumed that the main scale where 
$V({\bf r})$ change is essentially larger than the lattice 
constant. The equation (\ref{1}) holds in the bulk of crystal, 
but not at the interface. Therefore the certain boundary 
conditions have to be imposed to match the envelopes from  
both sides of the interface. 
The simplest boundary condition demands  the envelope and 
its derivatives  divided to the effective mass to be continued 
through the interface. The last 
condition ensures the probability flux conservation.

The problem becomes more complicated if the degeneration 
is present between the electron states from the different bands 
or valleys, i.e., if the equation $\varepsilon_n(k)=E$ has a 
few solutions. This is possible if any nonintrinsic symmetry 
element (a screw axis or a glide plane) is present in the 
crystal. The operator corresponded to this element moves the 
electron from one valley to another. This operator should 
commutate with the Hamiltonian, so that the electron energy  
does not change.  This is the reason for the degeneration. The 
nonintrinsic symmetry could not exist at the interface, so that 
the degeneration should be removed there. This means, that the 
certain relation between the wave functions from the 
different valleys should be fulfilled at the interface.  This  
relation implies the additional boundary condition, which has to 
be aroused from the degeneracy.

Let $M$ be the number of nonequivalent solutions of the equation 
$\varepsilon_n(k)=E$ for the lefthand semiconductor and $N$ be 
that for the righthand semiconductor. The solution of the 
Schr\"odinger equation is determined if the irradiation 
condition is specified at infinity. We can specify $M$ solutions 
for the lefthand semiconductor. This means that $M$ 
different plain waves could be incident on the contact from 
the left. In a similar manner $N$ solutions could be specified 
for the righthand semiconductor. Thus, the boundary conditions 
should be specified by the system of $M+N$  equations. 
If the effective mass approximation is applicable in both 
sides of the contact when it is sufficient to remain the quadratic 
term in the expansion of $\varepsilon_n(-i\nabla)$. 
 Then the boundary conditions 
should be represented by the system of $M+N$ linear equations for the 
envelopes and their derivatives at the boundary.

Let  us consider the contact between two materials 1 and 
2 (Fig.~1). The plains $S_1$ and $S_2$ are parallel to the 
boundary $z=0$ being removed from it so that  the electron wave 
functions  at these plains are Bloch. Nevertheless, the
distance between $S_1$ and $S_2$ is much more less then any 
characteristic size of the envelope. Thus, the space becomes 
divided on three regions: the left region ( $z<z_1$), the right 
region ( $z>z_2$), and the interface region ( $z_1<z<z_2$). Let 
$G({\bf r},\ {\bf r'})$ be the Green function of the boundary 
problem.  It obeys the equation 
\begin{equation} 
\label{2} 
\left[ 
-\Delta + 2m_e(U({\bf r}) -E)\right] G({\bf r},\ {\bf r'})=\delta 
({\bf r}-{\bf r'}).  
\end{equation} 
Here $U({\bf r})$ is the 
potential energy of the electron. It is the periodical function in 
the left and in the right regions, but it is not periodical in the 
interface region. 
 
Let  complete the plains $S_1$ and $S_2$ with the spheres at 
infinity as it is represented with the dotted line in Fig.~1. We 
can use the Green's theorem in order to connect the wave 
function at the surface $S_1$  to that  at the plain $S_2$
\widetext
\begin{mathletters}
\label{3}
\begin{equation}
\label{3a}
\FL
\psi({\bf r_2})=\int_{S_1} \left[G({\bf r_2},\ {\bf r_1}) 
\frac{\partial \psi({\bf r_1})}{\partial z_1} - \psi({\bf r_1})
\frac{\partial G({\bf r_2},\ {\bf r_1})}{\partial z_1} \right]
\, d^2{\bf r_1}.
\end{equation}
The wave function here is supposed to be vanished at the right 
infinity, so that the  equation (\ref{3a}) corresponds to the 
irradiation conditions where the plain wave is incident from the 
left.  We can apply also the Green's theorem to the surface $S_2$ to 
obtain the equation corresponded to the inverse irradiation 
conditions 
\begin{equation}
\label{3b}
\psi({\bf r_1})=\int_{S_2} \left[G({\bf r_1},\ {\bf r_2}) 
\frac{\partial \psi({\bf r_2})}{\partial z_2} - \psi({\bf r_2})
\frac{\partial G({\bf r_1},\ {\bf r_2})}{\partial z_2} \right]
\, d^2{\bf r_2}.
\end{equation}
\end{mathletters}
\narrowtext
We introduce the envelope wave functions as follows
\begin{equation}
\label{4}
\psi({\bf r})=\left\{ 
\begin{array}{ll}               
\sum\limits_{i=1}^M F_i^l({\bf r})\varphi_{k_i}^l({\bf r}),
& \mbox{if }z<z_1,  \\
\ &\ \\
\sum\limits_{i=1}^N F_i^r({\bf r})\varphi_{k_i}^r({\bf r}),
& \mbox{if }z>z_2.
\end{array}
\right.
\end{equation}
Here $\varphi_{k_i}^{l,\,r}({\bf r})$ are the Bloch wave 
function of the electron from the left and from the right 
regions and $F_i^{l,\,r}({\bf r})$ are the envelopes. The 
subscript $i$ enumerates here nonequivalent values of the wave 
vector ${\bf k_i}$ for that $\varepsilon({\bf k_i})=0$.
The envelopes $F_i^{l,\,r}({\bf r})$ are supposed to be smooth, 
so that the main size of their deviation considerably exceeds 
all other scales of the problem.

Substituting (\ref{4}) in (\ref{3}), multiplying the resulting 
equations by $\varphi_{{\bf k}_j}^{l,\, r*}$, and integrating them 
over the certain regions $\Omega_{1,\,2}$,   we obtain
\widetext
\begin{eqnarray}
\label{5}
&&\sum_{i=1}^M \int_{\Omega_1} F_i^l({\bf r_1})\varphi_{{\bf 
k}_i}^l({\bf r_1}) \varphi_{{\bf k}_j}^{l*}({\bf r_1})\,d^3{\bf r_1}= 
 \sum_{i=1}^N \int_{\Omega_1} \, \int_{S_1} \left\{ G({\bf 
r_1},\ {\bf r_2}) \frac{\partial}{\partial z_2} \left[F_i^r({\bf 
r_2})\varphi_{{\bf k}_i}^r({\bf r_2}) \right]-\right. 
\nonumber \\ 
&& \phantom{\sum_{i=1}^M \int_{\Omega_1} F_i^l({\bf r_1})\varphi_{{\bf 
k}_i}^l({\bf r_1}) \varphi_{{\bf k}_j}^{l*}({\bf r_1})\,d^3{\bf 
r_1}=}
\left. F_i^r({\bf 
r_2})\varphi_{{\bf k}_i}^r({\bf r_2}) \frac{\partial G({\bf r_1},\ 
{\bf r_2})}{\partial z_2} \right\} \varphi_{{\bf k}_j}^{l*}({\bf 
r_1})\,d^2 {\bf r_2}\,d^3 {\bf r_1},  \ \ j=1,\ \ldots,\ M \nonumber 
\\ \ \\ 
&&\sum_{i=1}^N \int_{\Omega_2} F_i^r({\bf r_2})\varphi_{{\bf 
k}_i}^r({\bf r_2}) \varphi_{{\bf k}_j}^{r*}({\bf r_2})\,d^3{\bf r_2}= 
 \sum_{i=1}^M \int_{\Omega_2} \, \int_{S_1} \left\{ G({\bf 
r_2},\ {\bf r_1}) \frac{\partial}{\partial z_1} \left[F_i^l({\bf 
r_1})\varphi_{{\bf k}_i}^l({\bf r_1}) \right]-\right. 
\nonumber \\ 
&&\phantom{\sum_{i=1}^N \int_{\Omega_2} F_i^r({\bf r_2})\varphi_{{\bf 
k}_i}^r({\bf r_2}) \varphi_{{\bf k}_j}^{r*}({\bf r_2})\,d^3{\bf 
r_2}=}
\left. F_i^l({\bf 
r_1})\varphi_{{\bf k}_i}^l({\bf r_1}) \frac{\partial G({\bf r_2},\ 
{\bf r_1})}{\partial z_1} \right\} \varphi_{{\bf k}_j}^{r*}({\bf 
r_2})\,d^2 {\bf r_1}\,d^3 {\bf r_2},\   \ j=1,\ \ldots,\ N. \nonumber 
\end{eqnarray}
\narrowtext
Here $\Omega_{1,2}$ are the vicinities of the points ${\bf r_1}$ and 
${\bf r_2}$. Their size is supposed to be large in 
comparison with the lattice constant, but it is small in comparison 
with the scale of envelope. In other words, $\Omega_{1,2}$ are large 
enough, so that the orthogonality conditions for the Bloch functions 
are satisfied, while the envelopes could be expanded in series at 
these regions. Equations (\ref{5}) specifies the $M+N$ boundary 
conditions for the envelopes. They are essentially nonlocal in 
agreement with \cite{FTT95,FTT97}.  If the effective mass 
approximation is applicable in both semiconductors,  then it is 
possible to retain only the quadratic term in (\ref{1}) and the 
linear term in expansion of the envelopes in  (\ref{5}). We obtain

\begin{eqnarray}
\label{6}
&&F_i^r+\tau_{i0}^{rx}\frac{\partial F_i^r}{\partial x}+
\tau_{i0}^{ry}\frac{\partial F_i^r}{\partial y}+
\tau_{i0}^{rz}\frac{\partial F_i^r}{\partial z} =
\sum_{j=1}^M\left( t_{ij}^lF_j^l+
\tau_{ij}^{lx}\frac{\partial F_j^l}{\partial x}+
\tau_{ij}^{ly}\frac{\partial F_j^l}{\partial y}+
\tau_{ij}^{lz}\frac{\partial F_j^l}{\partial z}\right)
\ \ \ i=1,\  \ldots,\ N, \nonumber \\
\ \\
&&F_i^l+\tau_{i0}^{lx}\frac{\partial F_i^l}{\partial x}+
\tau_{i0}^{ly}\frac{\partial F_i^l}{\partial y}+
\tau_{i0}^{lz}\frac{\partial F_i^l}{\partial z} =
\sum_{j=1}^N\left( t_{ij}^rF_j^r+
\tau_{ij}^{rx}\frac{\partial F_j^r}{\partial x}+
\tau_{ij}^{ry}\frac{\partial F_j^r}{\partial y}+
\tau_{ij}^{rz}\frac{\partial F_j^r}{\partial z}\right)
\ \ \ i=1,\  \ldots,\ M. \nonumber
\end{eqnarray}
There
\widetext
\begin{eqnarray}
\label{7}
&&\tau_{i0}^{rx}=\frac{1}{v_0}\int_{v_0}xu_{k_i}^r({\bf r})
u_{k_i}^{r*}({\bf r})\,d^3{\bf r},\ \ \                   
\tau_{i0}^{ry}=\frac{1}{v_0}\int_{v_0}yu_{k_i}^r({\bf r})
u_{k_i}^{r*}({\bf r})\,d^3{\bf r},\ \ \  
\tau_{i0}^{rz}=\frac{1}{v_0}\int_{v_0}zu_{k_i}^r({\bf r})
u_{k_i}^{r*}({\bf r})\,d^3{\bf r}, \nonumber \\
&&\tau_{i0}^{lx}=\frac{1}{v_0}\int_{v_0}xu_{k_i}^l({\bf r})
u_{k_i}^{l*}({\bf r})\,d^3{\bf r},\ \ \                   
\tau_{i0}^{ly}=\frac{1}{v_0}\int_{v_0}yu_{k_i}^l({\bf r})
u_{k_i}^{l*}({\bf r})\,d^3{\bf r},\ \ \  
\tau_{i0}^{lz}=\frac{1}{v_0}\int_{v_0}zu_{k_i}^l({\bf r})
u_{k_i}^{l*}({\bf r})\,d^3{\bf r}, \nonumber \\
&&t_{ij}^l=
\int_{\Omega_2} \, \int_{S_1} \left\{ G({\bf 
r_2},\ {\bf r_1}) \frac{\partial \varphi_{{\bf k}_j}^l
({\bf r_1})}{\partial z_1}   -
\varphi_{{\bf k}_j}^l({\bf r_1}) \frac{\partial G({\bf r_2},\ 
{\bf r_1})}{\partial z_1} \right\} \varphi_{{\bf k}_i}^{r*}({\bf 
r_2})\,d^2 {\bf r_1}\,d^3 {\bf r_2},  \\
&&\tau_{ij}^{lx}=
\int_{\Omega_2} \, \int_{S_1} (x_1-x_2)\left\{ G({\bf 
r_2},\ {\bf r_1}) \frac{\partial \varphi_{{\bf k}_j}^l
({\bf r_1})}{\partial z_1}   -
\varphi_{{\bf k}_j}^l({\bf r_1}) \frac{\partial G({\bf r_2},\ 
{\bf r_1})}{\partial z_1} \right\} \varphi_{{\bf k}_i}^{r*}({\bf 
r_2})\,d^2 {\bf r_1}\,d^3 {\bf r_2}, \nonumber \\
&&\tau_{ij}^{ly}=
\int_{\Omega_2} \, \int_{S_1} (y_1-y_2)\left\{ G({\bf 
r_2},\ {\bf r_1}) \frac{\partial \varphi_{{\bf k}_j}^l
({\bf r_1})}{\partial z_1}   -
\varphi_{{\bf k}_j}^l({\bf r_1}) \frac{\partial G({\bf r_2},\ 
{\bf r_1})}{\partial z_1} \right\} \varphi_{{\bf k}_i}^{r*}({\bf 
r_2})\,d^2 {\bf r_1}\,d^3 {\bf r_2}, \nonumber \\
&&\tau_{ij}^{lz}=
\int_{\Omega_2} \, \int_{S_1} \left\{ G({\bf 
r_2},\ {\bf r_1}) \frac{\partial }{\partial z_1}   
\left[z_1\varphi_{{\bf k}_j}^l ({\bf r_1})\right] -
z_1\varphi_{{\bf k}_j}^l ({\bf r_1})
\frac{\partial G({\bf r_2},\ 
{\bf r_1})}{\partial z_1} \right\} \varphi_{{\bf k}_i}^{r*}({\bf 
r_2})\,d^2 {\bf r_1}\,d^3 {\bf r_2}, \nonumber \\ 
&&t_{ij}^r=
\int_{\Omega_1} \, \int_{S_2} \left\{ G({\bf 
r_1},\ {\bf r_2}) \frac{\partial \varphi_{{\bf k}_j}^r
({\bf r_2})}{\partial z_2}   -
\varphi_{{\bf k}_j}^r({\bf r_2}) \frac{\partial G({\bf r_1},\ 
{\bf r_2})}{\partial z_2} \right\} \varphi_{{\bf k}_i}^{l*}({\bf 
r_1})\,d^2 {\bf r_2}\,d^3 {\bf r_1}, \nonumber \\
&&\tau_{ij}^{rx}=
\int_{\Omega_1} \, \int_{S_2}(x_1-x_2) \left\{ G({\bf 
r_1},\ {\bf r_2}) \frac{\partial \varphi_{{\bf k}_j}^r
({\bf r_2})}{\partial z_2}   -
\varphi_{{\bf k}_j}^r({\bf r_2}) \frac{\partial G({\bf r_1},\ 
{\bf r_2})}{\partial z_2} \right\} \varphi_{{\bf k}_i}^{l*}({\bf 
r_1})\,d^2 {\bf r_2}\,d^3 {\bf r_1}, \nonumber \\
&&\tau_{ij}^{ry}=
\int_{\Omega_1} \, \int_{S_2}(y_1-y_2) \left\{ G({\bf 
r_1},\ {\bf r_2}) \frac{\partial \varphi_{{\bf k}_j}^r
({\bf r_2})}{\partial z_2}   -
\varphi_{{\bf k}_j}^r({\bf r_2}) \frac{\partial G({\bf r_1},\ 
{\bf r_2})}{\partial z_2} \right\} \varphi_{{\bf k}_i}^{l*}({\bf 
r_1})\,d^2 {\bf r_2}\,d^3 {\bf r_1}, \nonumber \\
&&\tau_{ij}^{rz}=
\int_{\Omega_1} \, \int_{S_2} \left\{ G({\bf 
r_1},\ {\bf r_2}) \frac{\partial }{\partial z_2}   
\left[z_2\varphi_{{\bf k}_j}^r ({\bf r_2})\right] -
z_2\varphi_{{\bf k}_j}^r ({\bf r_2})
\frac{\partial G({\bf r_1},\ 
{\bf r_2})}{\partial z_2} \right\} \varphi_{{\bf k}_i}^{l*}({\bf 
r_1})\,d^2 {\bf r_2}\,d^3 {\bf r_1}, \nonumber 
\end{eqnarray}
\narrowtext
$v_0$ is the unit cell volume, and $u_{k_i}^{l,r}$ are the Bloch 
amplitudes. To determine $t_{ij}$ and $\tau_{ij}$ from Eq.~(\ref{7}) 
it is necessary  to obtain the Green function $G({\bf r_1},\ {\bf 
r_2})$ from Eq.~(\ref{2}). This is possible only if the potential 
$U({\bf r})$ is known at the interface. Usually it 
is not the case. Nevertheless, the certain general conclusions 
could be done. 

(a) If the points ${\bf k}_i$ are the particular points of the 
Brillouin zone, then the Bloch functions $\varphi_{{\bf 
k}_i}({\bf r})$ could be chosen as real. Therefore the 
parameters $t_{ij}$ and $\tau_{ij}$ are real as well. Moreover, 
$t_{ij}$ and $\tau_{ij}$, if they belong to the equivalent 
valleys are equal too. In general, they should be complex 
conjugated.

(b) From the integrands in Eq. (\ref{7}) it follows that 
$|\tau|\sim w|t|$ , where $w$ is the characteristic size of the 
interface  region ($w\sim a$ for a sharp interface and $w>a$ for 
a smooth one). Indeed, if we suppose the lattice constant to be 
vanished, then the potential $U({\bf r})$ becomes independent of $x$ 
and $y$. It follows from Eq.(\ref{2}) that $G({\bf r_2}, {\bf r_1})=
G(|{\bf r_2}- {\bf r_1}|)$ in this case. Then the integrals over the 
surfaces $S_1$ and $S_2$ in the expressions for $\tau^x$ and $\tau^y$ 
are vanished due to oddness of the integrands. In other words, the 
integrals (\ref{7}) are determined by the  regions, which  
size is of about the lattice constant.  As regards $\tau^z$, it is 
small due to existance of the value $z_2\sim w$ in the 
integrands.  

(c) It is possible to make use of the fact, that the boundary is the 
short-range potential in comparison with the mean size of the 
envelope.  That is why the scales of about $|{\bf r_2-r_1}|\sim 
w$ are significant in the integrands Eq.~(\ref{7}). Then it is 
possible to assume $E$ in Eq.(\ref{2}) to be equal to the energy 
of the relevant valley edge $E_b$. Such substitution 
\cite{Landau} leads to the appreciable error in Green function at the 
distances of about $(2m|E-E_b|)^{-1/2}\gg w$. In  other words the 
mean energy scale for $G({\bf r_1},\ {\bf r_2})$ and so for the 
parameters $t_{ij}$ and $\tau_{ij}$ is $|E-E_b|\sim (2mw^2)^{-1}$, i. 
e., the bandwidth. So that the parameters $t_{ij}$ and $\tau_{ij}$ 
could be regarded as independent of energy.

(d) It seems that it is possible to omit the terms 
with the derivatives in Eq. (\ref{6}). Really 
it is not the case.  Indeed, if the determinant consisted from 
the coefficients at the envelopes $F^{l,\,r}$ is not vanished, 
then omitting of $\tau F'$ leads to the trivial result 
$F^{l,\,r}(0)=0$. This means that the electron cannot even 
achieve the boundary. In fact $F^{l,\,r}$ is simply proportional to 
the small value in this case, $F(0)\sim w/\lambda$, so that  
each term in Eq. (\ref{6}) becomes of the same order. It was 
shown that the existence of this small parameter is really leads 
to the significant suppression (of the order of $(w/\lambda)^2$) 
of the interface transparency \cite{FTT95,FTT97}. This 
suppression disappears at the structure resonance, then the 
above-mentioned determinant is vanished, i.e., 
\begin{equation}
\label{8}
\left|
\begin{array}{cc}
E_{MM} & \|t_{ij}^r\| \\
\|t_{ij}^l\| & E_{NN}
\end{array}
\right| =0.
\end{equation}
Here $E_{MM}$ is the $M\times M$ identity matrix and 
$\|t_{ij}^{l,\,r}\|$ are the matrixes composed from the 
parameters (\ref{7}). In this case  Eq. (\ref{6}) could be 
represented as the  system of $M+N-1$ equations for the 
envelopes $F^{l,\,r}$ only and one equation for their 
derivatives $F'$. The latter equation does not contain the terms  
$t_{ij}F_j$ at the resonance and contains them with a small 
$t_{ij}$ nearby it.

There are $8MN+3(M+N)$ parameters in Eq. (\ref{7}). They are not 
independent because the probability flux should be continued 
through the interface.  The number of nonvanished parameters 
could be essentially reduced if the boundary is symmetrical.
This could be done by evolution of integrals (\ref{7}) with the 
group-theoretical method. It follows from Eq. (\ref{2}) that the 
Green function $G({\bf r}_1,\ {\bf r}_2)$ is invariant under the 
symmetry transformations connected with the symmetry of 
potential $U({\bf r})$.  This is possible if both boarded 
material as well as the boundary itself are symmetrical, i.e., if any 
axis or a plain is perpendicular to the interface $z=0$.
Then the certain parameters $t_{ij},\ \tau_{ij}$ in Eq.~(\ref{7})
becomes vanished.  This means that the  intervalley relation at 
the interface disappeared for such  valleys.

 It should be noted, that
the nonintristic symmetry elements, which have been the 
origin for  the intervalley degeneracy,  could not 
exist at the interface. Indeed, a screw axis or a glide plane 
 should distort the boundary shape or shift it in the distance of 
about the lattice constant. In other words, 
they  alter the potential $U({\bf r})$ at the interface. 
Nevertheless, if such distortions 
have not essentially change $G({\bf r_1},\ {\bf r_2})$, then the 
corresponding parameters $t_{ij},\ \tau_{ij}$ become small.
Two possibilities could be imagined in this way.  
Firstly, then the interface is smooth, i.e., if the periodic 
potential $U({\bf r})$ from the leftside material change to that from 
the rightside material smoothly on the size, which is much larger 
than the lattice constant ($w \gg a$).  In this case, the Bastard's 
boundary conditions are applicable for each pair of the equivalent 
valleys \cite{PR94,Burt}.  Secondly, if the bordered materials are 
not only crystallographically, but are chemically similar also.  
The pseudopotentials in both contacted materials are closely 
related in this case. So that the small deviation of the 
potential at the interface due to the nonintristic 
transformation should not appreciably affect $G({\bf r_1},\ {\bf 
r_2})$. Such situation is really occur in the heterojunctions 
GaAs/AlAs \cite{A61}.  It is difficult to imagine why the 
intervalley relation could disappear in more general case.

The equation (\ref{5}) allows to obtain the boundary conditions 
also if the effective mass approximation is unsufficial and the 
nonparabolicity of the band is essential. Then it is necessary 
to retain more terms in the expansion of Eq.~(\ref{5}). The 
received boundary conditions would contain the higher 
derivatives of the envelopes. The corresponding terms would be 
as small as $w/\lambda$.

\section{Light absorption at sharp boundary in indirect-band-gap 
semiconductor} 
\label{sec_3}

Let us consider the light absorption in the semiconductor, which band 
structure is presented in Fig.~2. We assume for simplicity that the 
valence band is not degenerate. There are two valleys in 
the conductive band: the central valley with the minima at the center 
of the Brillouin zone and the side valley at the edge  of it 
$(\pm \pi/a,\ 0,\ 0)$.   Suppose, also, that the symmetry of the 
crystal makes it possible to divide the variables, so that the 
problem becomes one-dimensional.

The probability for the photon to be absorbed is determined by the 
squared module of the matrix element
 
\begin{equation}
\label{9}
M=\int \psi_f^* H_{int} \psi_i \,dV
\end{equation}
Here
$$
H_{int}=-\frac{ie}{mc}A_0e^{-i(\omega t-\kappa_l z)}
\frac{\partial}{\partial z}, 
$$ 
$A_0$ is the vector potential of the light field, 
$\omega$ and $\kappa_l$ are the frequency and the wave number of 
light, the electric field of the light is supposed to be 
directed  along the $z$ axis, $ \psi_i$ and $\psi_f$  are the wave 
functions of the electron before the excitation (in the valence 
band and after it (in the conductive band); $\psi_i$  is a 
superposition of the incident wave and the divergent scattered 
wave, whereas $\psi_f$ is a superposition of the reflected 
wave and the convergent scattered wave \cite{Landau}.

\newpage

 We can write the wave functions as 
follows:
\widetext
\begin{eqnarray}
\label{10}
&&\psi_i=\left\{
\begin{array}{ll} 
u_v(z)e^{ipz}+Ru_v^*(z)e^{-ipz}, & z<0, \\
T_ve^{-\gamma_v z} & z>0,
\end{array}
\right.
\nonumber \\
\ \\
&&\psi_f=\left\{
\begin{array}{ll}
Au_{c0}(z)e^{\kappa z}+Bu_{cq}^*(z)
e^{i\left(q-\frac{\pi}{a}\right)z} +
u_{cq}(z)
e^{-i\left(q-\frac{\pi}{a}\right)z}, &  z<0, \\
T_ce^{-\gamma_c z}, & z>0.
\end{array}
\right.
\nonumber
\end{eqnarray}

Here $u_v(z)$, $u_{c0}$, and $u_{cq}$ are the Bloch amplitudes in the 
valent band, the central  and side valleyes of the conductive band, 
$p$, $\kappa$, and $q$ are the wave numbers  in these valleys. 
The wave function of the electron in the central valley is 
supposed to be decaying from the boundary. It should be noted 
that the decaying exponent might be not so large if the valley 
minima are close. The semiconductor occupies the region $z<0$, 
so that the wave functions are decaying then $z>0$.  The 
coefficients $R,\ T_v,\ A,\ B,\ T_c$ are determined by the 
boundary conditions (\ref{6}). We have 
\begin{eqnarray*} 
&&t_{11}^v(1+R)+ip\tau_{11}^v(1-R)=T_v(1-\gamma_v 
\tau_{10}^v), \\
&&T_v(t_{21}^v- \gamma_v \tau_{21}^v)= 
1+R+ip\tau_{20}^v(1-R), \\
&&(t_{11}^c+\kappa 
\tau_{11}^c)A+t_{12}^c(1+B)+iq\tau_{12}^c(B-1)= 
T_c(1-\gamma_c\tau_{10}^c), \\
&&A(1+\kappa\tau_{20}^c)=T_c(t_{21}^c-\gamma_c\tau_{21}^c), \\
&&1+B+iq\tau_{30}^c(B-1)=T_c(t_{31}^c-\gamma_c\tau_{31}^c).
\end{eqnarray*}
Hence
\begin{eqnarray}
\label{11}
&&R\simeq -1-\frac{2ip\left(t_{21}^v\tau_{11}^v-
\tau_{20}^v\right)}{t_{11}^vt_{21}^v-1-
ip\left(t_{21}^v\tau_{11}^v-\tau_{20}^v\right)}, \nonumber \\
&&A\simeq -\frac{2iqt_{21}^c\left(\tau_{12}^c-
t_{12}^c\tau_{30}^c\right)}
{1-t_{12}^c t_{31}^c-t_{11}^c t_{21}^c-iq
\left(t_{31}^c \tau_{12}^c+t_{11}^c t_{21}^c \tau_{30}^c -
\tau_{30}^c \right)}, \\
&&B\simeq -1 -\frac{2iq\left(t_{31}^c\tau_{12}^c+
t_{11}^ct_{21}^c\tau_{30}^c-\tau_{30}^c\right)}
{1-t_{12}^c t_{31}^c-t_{11}^c t_{21}^c-iq
\left(t_{31}^c \tau_{12}^c+t_{11}^c t_{21}^c \tau_{30}^c -
\tau_{30}^c \right)}. \nonumber
\end{eqnarray}
\narrowtext
Here $t_{ij}^{c,\,v},\ \tau_{ij}^{c,\,v}$ are the boundary 
parameters determined by Eq.(\ref{7}). It is assumed that 
$|k\tau|\ll1$ and $|q\tau|\ll1$.  The structure resonance 
conditions (\ref{8}) in this case means that the denominators   
in Eq.~(\ref{11}) have vanished.  It is clear from Eq.(\ref{11}) 
that $|R|=1,\ |B|=1$. Thus apart from the resonance $R=-1$ and 
$B=-1$, whereas the conversion degree $A\sim a/\lambda$. If the 
resonance occurs for a valence band, then $R=1$.  For the 
resonance in conductive band $B=1$ and $A=2t_{21}/t_{31}$.

Let us determine the matrix element (\ref{9}) with the wave 
functions (\ref{10}, \ref{11}):
\begin{eqnarray}
\label{12}
&&M=-\frac{ie}{mc}A_0Se^{i(\varepsilon_c-\varepsilon_v-\omega)t}
{\cal P}_{vc},\ \mbox{then} \nonumber \\
&&{\cal P}_{vc}=\int_{-\infty}^0 \left[A^*u_{c0}(z)e^{\kappa z}+
B^*u_{cq}(z)e^{-i\left(q-\frac{\pi}{a}\right)z}+\right. \\
&&\left. u_{cq}^*(z)e^{i\left(q-\frac{\pi}{a}\right)z}\right] 
\frac{\partial}{\partial z} 
\left[u_v(z)e^{ipz}+Ru_v^*e^{-ipz}\right] \, dz. \nonumber
\end{eqnarray}
The wave number of light is assumes to be vanished in 
Eq.(\ref{12}). We could take advantage of the periodicity of 
Bloch functions to evaluate the integral (\ref{12}), then
\begin{equation}
\label{13}
{\cal P}_{vc}=A^* P_{c}
\left(\frac{1}{\kappa+ip}+\frac{R}{\kappa-ip}\right) +
\frac{a}{2}P_{s}(1+B^*)(1+R).
\end{equation}
Here 
$$
P_{c}=\frac{1}{a}\int u_{c0}^*\frac{\partial 
u_v}{\partial z}\,dz,\ \ \ 
P_{s}=\frac{1}{a}\int u_{cq}^*\frac{\partial 
u_v}{\partial z}\,dz.
$$

Two terms in Eq.(\ref{13}) could be interpreted as follows. The 
first one corresponds to the exitation of the electron to the 
surface state of the central valley  subsequented by the 
conversion to the side valley. It could be prevailing if the valleys 
minima are close,  so that  $\kappa a\ll 1$. The second term in 
Eq.~(\ref{13}) correspondes to the immediate  transition of 
the electron to the side valley at the interface (Fig.~2).

The absorption constant should be expressed as follows:
\begin{equation}
\label{14}
\alpha=\frac{2}{(2\pi)^2 Nv}\int |{\cal P}_{vc}|^2 
\delta(\varepsilon_c- \varepsilon_v-\omega)\,d^3{\bf k} 
\end{equation}
The magnitude of $\alpha$ substantially depends on  
presence  or  absence of the structure resonance in any 
band.
 If the 
resonance is absent in either band then $A\sim iq\tau,\ 1+R\sim 
ip\tau$, and $1+B^*\sim iq\tau$. Then from (\ref{13}, \ref{14}) 
 ${\cal P}_{vc} \propto kq$ and so
\begin{mathletters}
\label{15}
\begin{equation}
\label{15a}
\alpha \propto (\omega-\varepsilon_g)^4 \left( \frac{\pi a}{L} 
\right)^6 \left[|P_{s}|^2+\beta \frac{P_{c}^2}{(\kappa 
a)^2}\right], 
\end{equation} 
there $L$ is the absorption length, $\beta$ is the 
dimensionless parameter arisen from the first term in expression 
(\ref{13}). 
If 
the resonance takes place in any band, then ${\cal P}_{vc} 
\propto k,\ q$, and 
\begin{equation} 
\label{15b} 
\alpha\propto (\omega-\varepsilon_g)^3 \left( \frac{\pi a}{L} 
\right)^4 \left[|P_{s}|^2+\beta \frac{P_{c}^2}{(\kappa 
a)^2}\right].  
\end{equation} 
For the resonant case in both bands 
\begin{equation}
\label{15c}
\alpha\propto(\omega-\varepsilon_g)^2 \left( \frac{\pi a}{L} 
\right)^2 \left[|P_{s}|^2+\beta \frac{P_{c}^2}{(\kappa 
a)^2}\right].
\end{equation}
\end{mathletters}
It is clear that the absorption magnitude as well as its 
frequency dependence at the edge are determined by the 
conditions at the boundary.

\section{Discussion}
The boundary conditions for the envelope wave function proposed 
in this paper are applicable for the contacts of the materials 
with rather different band structures. They are suitable also if 
the degeneracy degrees in both bordered crystals are 
different.  
The present boundary conditions do not assume the band offsets 
to be small at the interface, i.e., they are appropriate not 
only for the heterojunctions, but also at the boundaries metal 
--semiconductor, metal--insulator,  inorganic -- organic 
contacts, and at the  crystalline surface. It is shown that 
the boundary conditions  are defined as  array 
of the linear equations for the envelopes and their 
derivatives. The coefficients at the derivatives there are shown 
 to be small.   However, they are important then the 
 transport through the boundary as well as the contact phenomena 
are considered.

It is easy to understand why the coefficients at the 
derivatives are so small. At first, let consider the ordinary 
one-dimensional quantum mechanical problem for the electron in 
the potential wall. The $same$ Schr\"odinger equation holds at 
both sides of the wall. So that, in order to define the solution 
of this equation it is necessary to define the wave function  
 in any two points . Would we choose the distance 
between these points to be infinitesimal, then the wave function 
and its derivative become given at the point.  This is the 
ordinary matching condition for the wave function. However, 
this is not the case for the envelope wave functions, which 
obeys the $different$ equations (\ref{1}) at each side of the 
interface.  Moreover, the equation (\ref{1}) does not 
hold in the interface region. The Schr\"odinger 
equation with the proper interface  potential has to 
be considered in this region. Its solution defines the 
wave functions, in which the lattice constant (it might be 
different at each side of the contact) is the 
particular parameter. As it has been shown in the exact 
solveble models \cite{FTT95,FTT97}, this leads to the 
nonlocality of the boundary conditions. This is true in the 
 general case as well. This means, that it is 
impossible to match the envelopes at the arbitrary points.   
So that the envelopes from each 
side have to be matched not at the $same$ site, but at the 
$different$ sites.  The positions of these sites ${\bf r_i}$ 
could be obtained from Eq.~(\ref{6}), if we suppose there 
$F({\bf r_i})=F(0)+{\bf \tau \nabla}F(0)$ for each envelope or 
expand the envelopes in Eq.~(\ref{5}) in these particular points, 
where $\nabla F({\bf r_i})=0$.  The distances between these 
sites could not be chosen arbitrary, so that the simple relation  
for the derivatives is no longer exist. Nevertheless, it would 
appear reasonable that these distances should be of the order of 
the lattice constant $a$.  Then $\tau \sim a$.  These 
qualitative arguments are confirmed by the results of this 
paper.

It is shown that the structure resonance occurs not only in the 
simplest models \cite{FTT95,FTT97}, but in the general case as 
well.  At first glance Eq.~(\ref{8}) is not practicable. Really, 
it is not the case. To understand the physical meaning of the 
structure resonance let us consider the electron scattering at 
the boundary.  In the simplest cases of the one-valley and 
two-valley semiconductors the wave functions could be written  
as $\psi_i$ and $\psi_f$ in Eq.~(\ref{10}). The reflection 
coefficients ($R$ and $B$) for both cases are determined by 
Eq.~(\ref{11}).  The resonant condition  
 (\ref{8}) is satisfied if the real parts of the 
denominators Eq.~(\ref{11}) are vanished.  On the other hand, 
the poles of the reflection coefficients determine the positions 
of the surface levels. Thus, the structure resonance condition 
(\ref{8}) means that the energy position of a surface 
level is close to the edge of the relevant band. 
As a rule such levels do exist at the 
semiconductor surface and they are responsible for the pinning 
of the Fermi level at the interface.  If so, then the electron 
with Fermi energy in bulk should be in resonance then crossing 
the interface. 

It is known, that the large amount of the  surface 
states are existing in the gap of the amorphous semiconductors. 
So, the structure resonance  could be  expected at the 
interface with an amorfous semiconductor.  This might manifest 
itself, in particular, as the difference of the optical spectra 
of the same semiconductor nanocrystal in the different (amorfous 
and crystalline) matrices \cite{120}. The effect might be 
associated with the difference of the quantization conditions for 
an electron at the resonance (at the interface with the amorfous 
matrix) and beyond from it. The  envelope wave 
function of the electron could be assumed to be vanished at the 
boundary in the 
absence of the resonance.
This is  the quantization condition in this case. 
However, it is not the case at the resonance.  
Then the crystallite effective size should 
increase, so that the electron could spend more time in the 
matrix. The possibility of the intervalley conversion also should 
 affect the electron spectrum.

It is probable that the suppression of the boundary transparency 
apart from the structure resonance is responsible for the blue 
shift in the absorption spectra in TiO$_2$  observed in 
Ref.~\cite{Kavan}. The blue shift there is associated with the 
electron confinement in the small (about the few nm) 
crystallites.  This crystallites  compose the thick (of the 
order the hundred nm) film, which has been prepared by the anode 
oxidative method. It is important that the conductance of the 
layer should be large enouph to ensure the film growth. Thus, 
the layer should be conductive in order to the film growth, 
while the electrons sould be confined in order to the blue shift 
becomes observeble.  To understand the phenomenon it should be 
emphasized that the boundary transparency is suppressed only for 
the long-wavelength electrons.  The electrons moved by the light 
to the bottom of the conductive band are long-wavelength 
($\lambda \sim L$, where $L$ is the cristallite size). So that 
the transparency factor $(a/\lambda)^2$ is small for them.  This 
is not the case for the electrons  moved to the conductive band 
from the outside. Thus, the quantum size effect concerns only 
the electrons excited by light, whereas the electrons with 
$\lambda \sim a$ maintaine the current, which is necessary for 
the film growth. 

It is shown that the boundary conditions at the sharp interface 
could be described by the  parameters, which are 
independent of energy. Their number seems to be large for the 
many-valley semiconductors. Nevertheless, it could be 
significantly reduced if the bordered materials as well as the 
boundary itself are symmetrical.  The group-thoretic method 
may be used for this \cite{Kisin}. The expressions for the 
parameters (\ref{11}) are appropriate for this purpose. 
 The boundary conditions parameters could be 
either calculated from the microscopic models or measured in the 
experiments. It is important, that the impurities as well as the 
structure imperfections would they exist at the interface should 
affect the parameters. Thus, the magnitude of the parameters for 
the interface should depend on the technology of its 
preparation.  The latter factor hardly could be taken into 
account in the numerical simulations, but it should manifest 
itself in the experiments.  The possible experimental ways for 
the determination of the boundary conditions parameters will be 
considered in the separate publication.

The boundary conditions obtained in the paper are applied for the 
investigation of the light absorption at the surface of the indirect-band-gap 
semiconductor. This is of particular importance in the restricted 
semiconductors where the share of the surface atoms is sufficiently 
large. It is shown that the enhancement of the effect has to take 
place.  Firstly, it happens owing to the immediate electron 
transition to the side valley at the interface where the 
nonconservation of the electron momentum is possible.  Secondly, 
it happens due to the conversion of the electron from the surface 
state at the central valley to the side one. The latter mechanism is 
essential if the energy positions of the valley minima ($\delta E$) 
are close.  Such a situation occurs in Ge and in some semiconductors 
of A$_{\rm III}$B$_{\rm V}$ group, there $\delta E \sim 0.1$\,eV, so 
that the effect has to increase in $(\kappa a)^{-2} \sim 100$ times. 
Moreover, the valley minima possibly becomes closer in the 
nanocrystallites of the porous semiconductors (TiO$_2$, p-Si).

It is shown that the absorption value $\alpha$ as well as its dependence on 
the photon energy at the absorption edge are sensitive  to  the  
structure resonance  whether it is present at the interface. The 
small parameter, to which the absorption is proportional, varied 
 according to Eq.(\ref{15}) from $(\pi a/L)^6$, if the resonance is 
absent in any band to $(\pi a/L)^2$ then it is present in both bands. 
It is easy to understand the reason. The wave function of the 
electron is small at the interface in the absent of the structure 
resonance, i.e., the electron spends too little time at the 
interface, where  the light absorption without the momentum 
conservation is possible.

Let us estimate the minimal size of the crystallite, in which the mechanism of 
the light absorption proposed in Sec.\,IV is essential. We have to compare the 
small parameter aroused in Eq.(\ref{15}) to that aroused in the phonon assisted 
light absorption. It is known, that the contribution of the latter mechanism 
to $\alpha$ is small in comparison with the contribution of the 
direct vertical interband electron transition as 10$^{-2}$ -- 
10$^{-3}$.  $(\pi a/L)^6\sim 10^{-3}$ then $L\sim 10a$, i.e., for the 
crystallite size of about 30 -- 50\,\AA . It has a vary rapid  
increase   then $L$ becomes less. It is quite possible, that the 
enhancement of the light absorption in the nanocrystalls, which has 
been observed in the experiments \cite{186} could be connected with 
the mechanism considered in the paper.

The  structure resonance in any or in  both bands leads to the essential
enhancement of the absorption. The latter could be possible in the amorfous 
semiconductors, where the density of the surface levels at both band extrema 
is high. The value $\alpha$ is determined by Eq.(\ref{15c}) in this 
case, so that $(\pi a/L)^2\sim 10^{-3}$,  then $L\sim 100a$. The 
significant (in 100 times) enhancement of the light absorption has 
been observed in the a-Si  layers, which thickness is about 
1$\mu$m \cite{Beck}.

It follows from Eq.~(\ref{15}), that absorption should rapidly 
increase when the crystallite size becomes less. This means, 
that in the porous materials where the strong dispersion of the 
crystallite size occurs, the main contribution to the absorption 
comes from the smallest of them. Thus, the estimation of the 
 crystallites size from the blue shift in the absorption spectra 
is really determines its lower bound.

\section{Summary}
The boundary conditions for the envelope wave function, which are 
appropriate for the contacts of rather different  materials have 
been proposed. It is shown that the boundary conditions are 
determined by the certain array of the parameters, depended only on 
the interface structure and the bordered materials themself. 
These parameters essentially identify all the properties of the 
interface, which can be described in terms of the envelope wave 
functions.  The values of these parameters could be found numerically 
from the appropriate microscopeing models or measured in experiments.  

It is shown that the electron behavior at the interface 
significantly depends on the structure resonance, whether it is  
occurring at the interface [Eq.~(\ref{8})]. The absence of the 
resonance means that the electron wave function is small at the 
interface, so that all the physical phenomena governed by the 
boundary becomes suppressed.

The light absorption at the indirect-band-gap semiconductor interface 
is investigated. It is shown that the possibility of the 
nonconservation of the electron momentum at the interface made the 
significant enhanacement of the absorption in the crystallites, 
which are as small as 30 -- 50\,\AA. The essential influence of 
the structure resonance at the interface on the absorption has 
been demonstrated.

\acknowledgments
Author would like to thank Dr.~D.~Romanov, Dr.~E.~Baskin and 
Dr.~M.~Entin for helpful discussions. This work was supported by 
the Russian Foundation for the Fundamental Inverstigarions Grant 
No.~96-02-19028, Russian Universities Foundation Grant 
No.~95-0-7.2-151 and 
the Swiss National Science Foundation Project 
No.~7SUPJ048573.

\begin{figure}
\caption{The Cartesian coordinate system adopted in the 
problem. Two materials 1 and 2 are separated by the plain $z=0$. The 
dotted lines indicate the extensions of the plains $S_1$ and $S_2$ at 
infinity.}  
\label{prbbrag1} 
\end{figure}
\begin{figure}
\caption{The two ways of the light absorption at the 
indirect-band-gap semiconductor surface: the immediate electron 
transition to the side valley $P_s$ and  the  vertical 
transition $P_c$ followed by the conversion to the side valley 
(dotted arrow).} 
\label{prbbrag2} 
\end{figure}
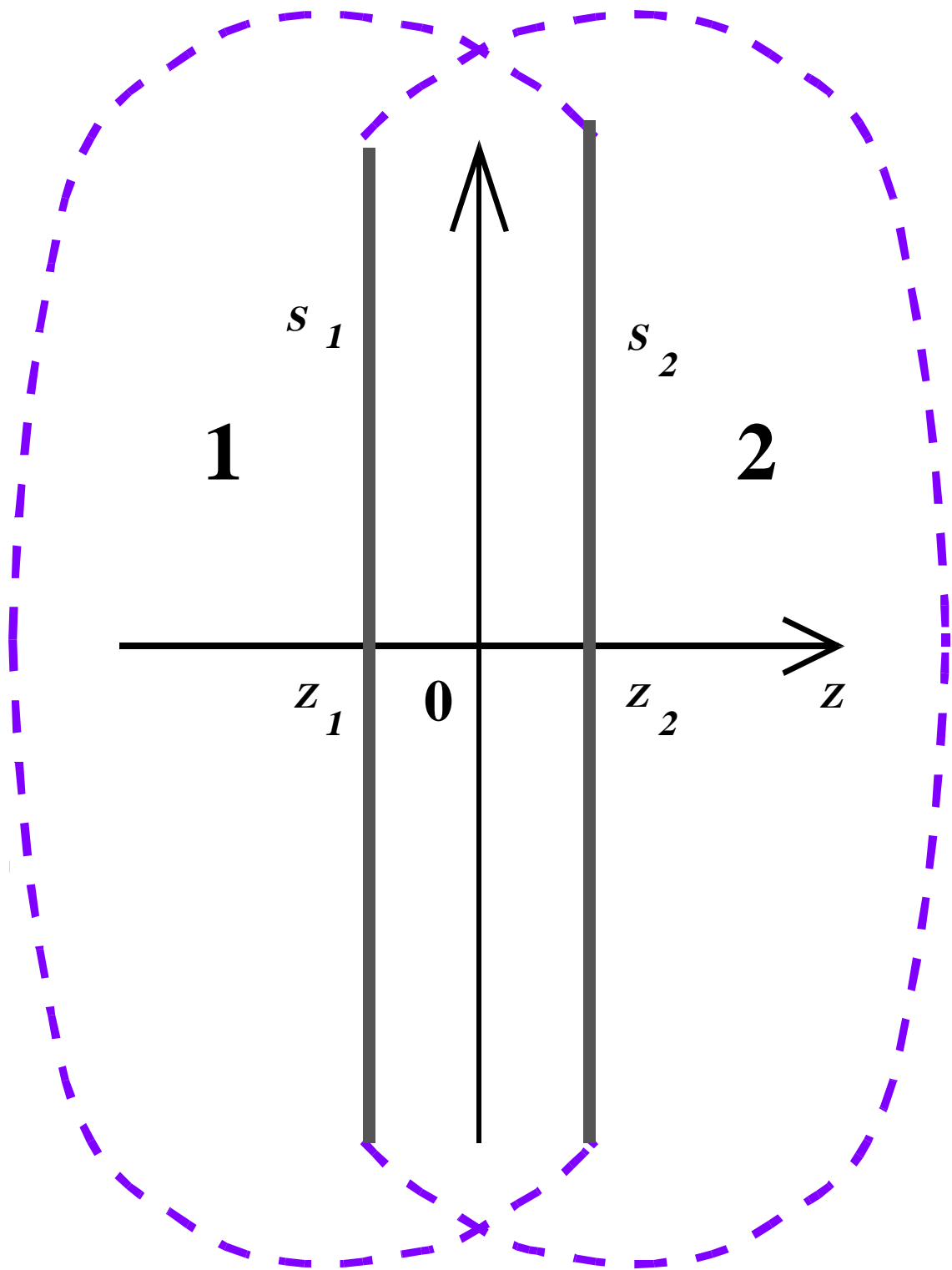
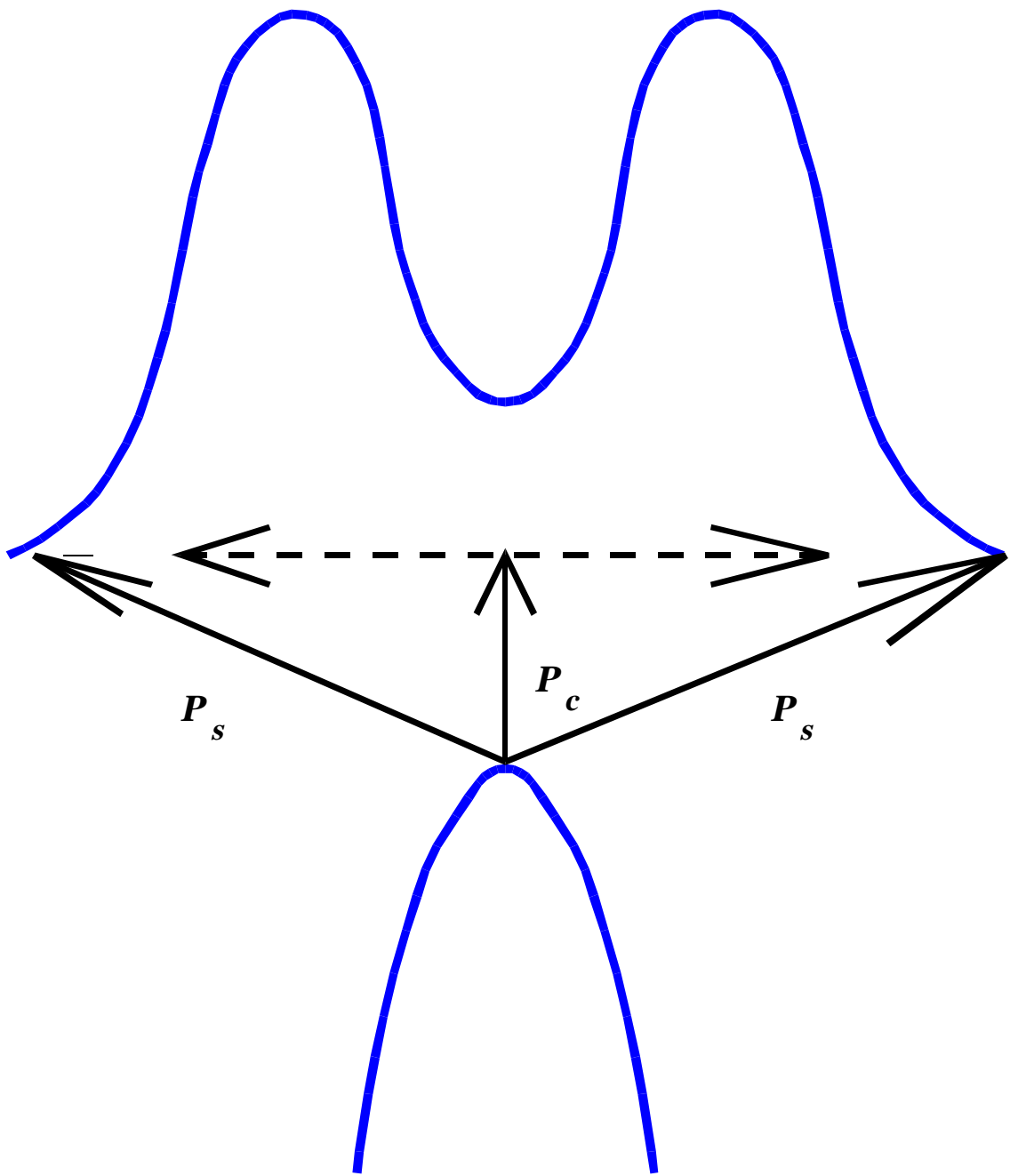
\end{document}